\documentclass[twocolumn,showpacs,preprintnumbers,superscriptaddress,prb,amsmath,amssymb]{revtex4-1}

\usepackage{graphicx}
\usepackage{dcolumn}
\usepackage{bm}
\usepackage{graphicx}
\usepackage{color}
\usepackage{ulem}
\usepackage{gensymb}
\usepackage{braket}
\usepackage{amsmath}
\usepackage[percent]{overpic}

\begin{document}

\title{Lattice distortion in the spin-orbital entangled state in \textit{R}VO$_3$ perovskites}

\author{J.-Q. Yan}
\email{yanj@ornl.gov}
\affiliation{Materials Science and Technology Division, Oak Ridge National Laboratory, Oak Ridge, Tennessee 37831, USA}

\author{W. Tian}
\affiliation{Neutron Scattering Division, Oak Ridge National Laboratory, Oak Ridge, Tennessee 37831, USA}

\author{H. B. Cao}
\affiliation{Neutron Scattering Division, Oak Ridge National Laboratory, Oak Ridge, Tennessee 37831, USA}

\author{S. Chi}
\affiliation{Neutron Scattering Division, Oak Ridge National Laboratory, Oak Ridge, Tennessee 37831, USA}

\author{F. Ye}
\affiliation{Neutron Scattering Division, Oak Ridge National Laboratory, Oak Ridge, Tennessee 37831, USA}

\author{A. Llobet}
\affiliation{P-23, Neutron Science and Technology, Los Alamos National Laboratory, Los Alamos, New Mexico 87545, USA}

\author{Q. Chen}
\affiliation{Department of Physics and Astronomy, University of Tennessee, Knoxville, Tennessee 37996, USA}

\author{J. Ma}
\affiliation{Key Laboratory of Artificial Structures and Quantum Control, School of Physics and Astronomy, Shanghai Jiao Tong University, Shanghai 200240, P. R. China}

\author{Y. Ren}
\affiliation{ X-ray Science Division, Argonne National Laboratory, Argonne, IL 60439, USA}

\author{J.-G. Cheng}
\affiliation{Beijing National Laboratory for Condensed Matter Physics, and Institute of Physics, Chinese Academy of Sciences,
Beijing 100190, P. R. China}

\author{J.-S. Zhou}
\affiliation{Materials Science and Engineering Program, University of Texas at Austin, Austin, Texas 78712, USA}

\author{M. A. McGuire}
\affiliation{Materials Science and Technology Division, Oak Ridge National Laboratory, Oak Ridge, Tennessee 37831, USA}

\author{R. J. McQueeney}
\affiliation{Ames Laboratory and Department of Physics and Astronomy, Iowa State University, Ames, Iowa 50011, USA}
\date{\today}

\begin{abstract}

We report a thorough study of Y$_{0.7}$La$_{0.3}$VO$_3$  single crystals by measuring magnetic properties, specific heat, thermal conductivity, x-ray and neutron diffraction with the motivation of revealing the lattice response to the spin-orbital entanglement in \textit{R}VO$_3$. Upon cooling from room temperature, the orbitally disordered paramagnetic state  changes around T*$\sim$220\,K to spin-orbital entangled state which is then followed by a transition at T$_N$=116\,K to C-type orbital ordered (OO) and G-type antiferromagnetic ordered (AF) ground state. In the temperature interval T$_N<T<T^*$, the VO$_{6/2}$ octahedra have two comparable in-plane V-O bonds which are longer than the out-of-plane V-O1 bond. This local structural distortion supports the spin-orbital entanglement of partially filled and degenerate yz/zx orbitals. However, this distortion is incompatible with the steric octahedral site distortion intrinsic to orthorhombic perovskites. Their competition induces a second order transition from the spin-orbital entangled state to C-OO/G-AF ground state where the long range OO suppresses the spin-orbital entanglement. Our analysis suggests that the spin-orbital entangled state and G-OO are comparable in energy and compete with each other. Rare earth site disorder favors the spin-orbital entanglement rather than a cooperative Jahn-Teller distortion. The results also indicate for LaVO$_3$ a C-OO/G-AF state in T$_t$\,$\leq$\,T\,$\leq$T$_N$ and an orbital flipping transition at T$_t$.

\end{abstract}

\maketitle

\section{Introduction}
\textit{R}VO$_3$ (\textit{R}=rare earth and Y) perovskites, in which V$^{3+}$ has an electronic configuration of t$_{2g}^2$, show a series of spin and orbital ordering states below room temperature and have been an ideal material playground for the study of complex coupling between spin, orbital, and lattice degrees of freedom  of t$_{2g}$ electrons.\cite{miyasaka2003spin} Despite the fact that the spin and orbital ordering patterns of \textit{R}VO$_3$ perovskites can be well understood following Goodenough-Kanamori rules,\cite{goodenough1955theory,kanamori1959superexchange} the quantum effects are widely believed to be important for understanding the spin and orbital ordered ground states due to a weak Jahn-Teller coupling of  t$_{2g}$ electrons. It has been under hot debate whether the classical Jahn-Teller orbital physics, or quantum orbital fluctuations, or a close interplay between them better account for the unusual behavior of t$_{2g}$  electrons in vanadate perovskites. \cite{blake2009competition,oles2007one,motome2003one,de2007orbital,khaliullin2001spin,fang2004quantum,oles2006spin,yan2004unusually,solovyev2006lattice,reul2012probing, skoulatos2015jahn}  In addition to the competition between different spin/orbital ordered ground states, spin-orbital entanglement from Kugel-Khomskii(KK)-type superexchange interactions\cite{ki1982jahn} competes with the long-range ordered states and should be considered when understanding the orbital physics in \textit{R}VO$_3$  perovskites.\cite{oles2012fingerprints}

\begin{figure} \centering \includegraphics [width = 0.38\textwidth] {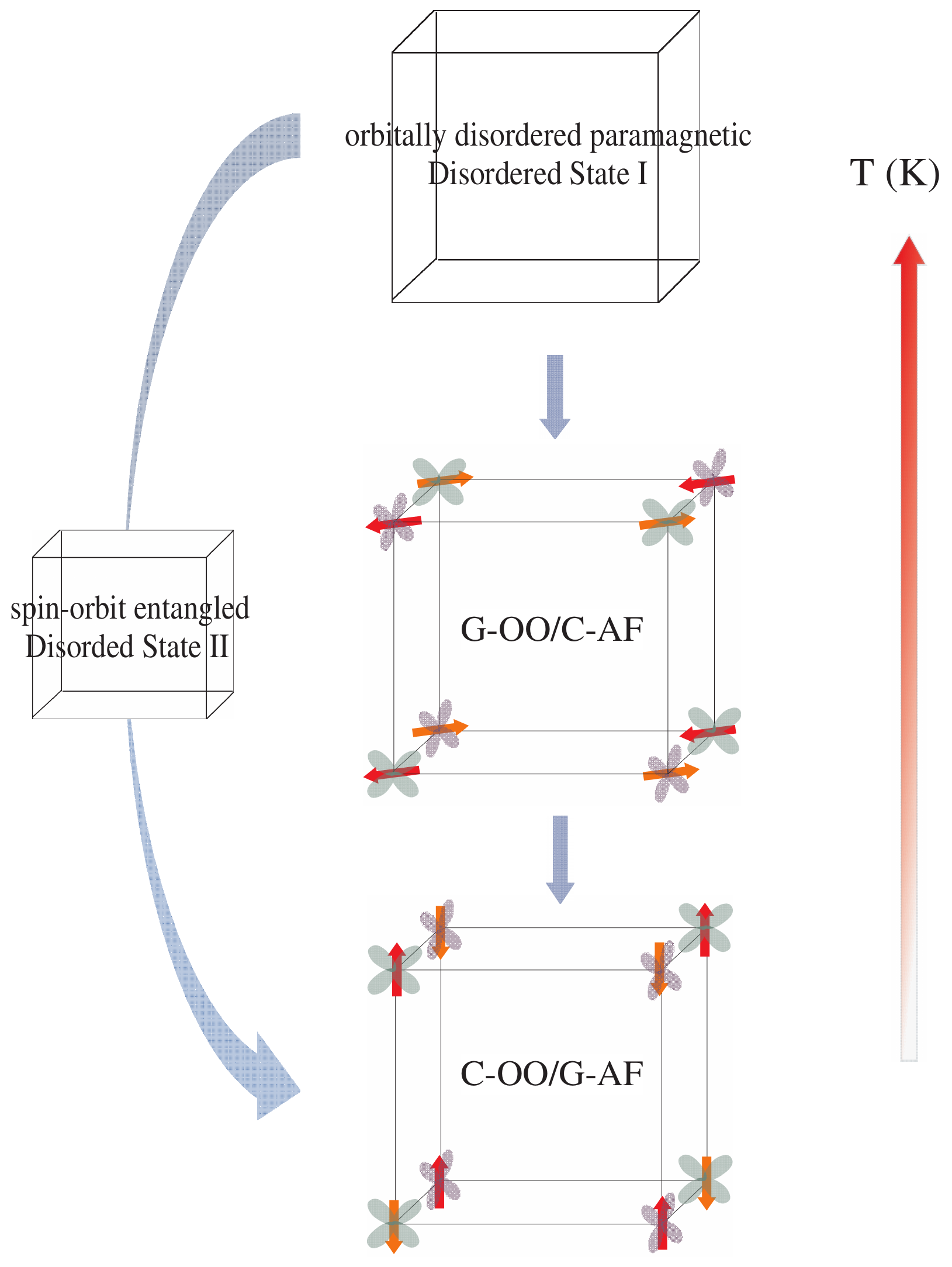}
\caption{(color online) Two different routes to the C-OO/G-AF state from the orbitally disordered paramagnetic state in \textit{R}VO$_3$. OO: orbital ordering. AF: antiferromagnetic. Only yz and zx orbitals are plotted. xy orbital is occupied on every V site. yz and zx orbitals alternate in \textit{ab}-plane. G-type: out-of phase order along \textit{c}-axis. C-type: in-phase arrangement along \textit{c}-axis. It should be noted that orbital order into G-type occurs at a higher temperature than magnetic order into C-type for \textit{R}=Pr,...,Lu and Y. }
\label{OO-1}
\end{figure}

With the largest rare earth ions among \textit{R}VO$_3$ members, the pseudocubic LaVO$_3$ has been extensively studied to understand the effect of orbital fluctuations on the magnetic ground state and anisotropic optical properties.\cite{khaliullin2001spin,motome2003one,de2007orbital,fang2004quantum,miyasaka2005one,fang2003anisotropic,kim2017signatures} A detailed experimental study\cite{zhou2008frustrated} on LaVO$_3$ single crystal suggests a spin-orbital entangled state in the temperature range T$_N$=143\,K$<$T$<$T*$\approx$200\,K. Despite careful neutron and synchrotron diffraction studies,\cite{zhou2008frustrated,ren2003orbital,tung2008spin,bordet1993structural,seim1998non} the octahedral distortion associated with the formation of the spin-orbital entangled state is unknown. The challenge comes from the fact that all three V-O bond lengths in LaVO$_3$ are comparable to each other above T$_N$. In LaVO$_3$, a first order structural transition takes place at T$_t$=141\,K, just below T$_N$=143\,K.\cite{ren2003orbital,tung2008spin} The origin for this structural transition is unknown. As a consequence, the spin and orbital patterns in the narrow 2\,K temperature window between T$_N$ and T$_t$ are under debate. It is worth mentioning that previous theoretical studies mainly investigate how the quantum fluctuations induce the G-OO/C-AF ground state without considering the details and importance of possible spin and orbital ordering in that 2\,K range.

A systematic study\cite{yan2011spin} on the spin and orbital ordering of Y$_{1-x}$La$_x$VO$_3$ shows that the spin-orbital entangled state (labelled as disordered state II in Ref[\citenum{yan2011spin}]) exists in a wide compositional range 0.2$<$x$\leq$1. The comparative study between Y$_{1-x}$La$_x$VO$_3$  and \textit{R}VO$_3$ suggests two different routes, as shown in Fig.\,\ref{OO-1}, to the C-OO/G-AF state from the orbitally disordered paramagnetic state: either via the G-OO/C-AF state or the spin-orbital entangled disordered state II. The former is observed in \textit{R}VO$_3$ (\textit{R}=Dy,...,Lu) and Y$_{1-x}$La$_x$VO$_3$ (0$\leq\,x\,<$0.20). Upon cooling from room temperature, these vanadate perovskites show a G-type orbital order around 200\,K, a magnetic order around 100\,K, and then an orbital flipping transition around 70\,K to the C-OO/G-AF state. The latter route to the C-OO/G-AF state occurs in \textit{R}VO$_3$ (\textit{R}=Y and Tb) under high pressure\cite{bizen2008orbital}, Y$_{1-x}$La$_x$VO$_3$ (0.20$<$x$\leq$1.0), and Eu$_{1-x}$(La$_{0.254}$Y$_{0.746}$)$_x$VO$_3$ (x$>$0.9)\cite{fukuta2011effects}. While the evolution with temperature of the nuclear structure in the former route has been extensively studied,\cite{blake2002neutron,blake2009competition,reehuis2006neutron,marquina2005lattice,reehuis2011structural,martinez2008evolution,munoz2003structural} a detailed investigation of the lattice response in the latter route is still absent.\cite{fukuta2011effects,bizen2008orbital}

In this paper, we present a thorough study of Y$_{0.7}$La$_{0.3}$VO$_3$  single crystals by a variety of techniques including magnetic measurements, specific heat, thermal conductivity, x-ray and neutron diffraction. Compared to the pseudocubic LaVO$_3$ with three nearly identical V-O bonds, Y$_{0.7}$La$_{0.3}$VO$_3$ has a larger GdFeO$_3$ distortion and has, at room temperature, one long, one medium, and one short V-O bond that could be well resolved with neutron powder diffraction measurements. Therefore, Y$_{0.7}$La$_{0.3}$VO$_3$ provides an ideal material platform to study the local lattice change in the spin-orbital entangled state. Our structural study shows that the splitting between the in-plane V-O bonds becomes smaller upon cooling in the temperature range T$_N$\,$\leq$\,T\,$\leq$300\,K. This provides strong evidence for the spin-orbital entanglement of degenerate yz and zx orbitals in the disordered state II. This structural response is incompatible with the octahedral site distortion intrinsic to orthorhombic perovskites; and their competition induces a second order transition from the spin-orbital entangled state to C-OO/G-AF ground state. Our analysis suggests that the spin-orbital entangled state and G-OO are comparable in energy and rare earth site disorder favors the spin-orbital entanglement rather than a cooperative Jahn-Teller distortion. The results also indicate for LaVO$_3$ a C-OO/G-AF state in the narrow range from T$_t$\,$\leq$\,T\,$\leq$T$_N$ followed by an orbital flipping transition at T$_t$.

\section{Experimental Details}

Y$_{0.70}$La$_{0.30}$VO$_3$ single crystal was grown using an optical floating zone furnace in a procedure described elsewhere.\cite{yan2004unusually}. The room-temperature x-ray diffraction experiment on powder from pulverized single crystals confirmed single phase and the diffraction pattern could be indexed in Pbnm symmetry. A large thermoelectric power ($>$550 $\mu$V/K) at room temperature implies an essentially stoichiometric oxygen content. Elemental analysis confirmed the Y/La ratio of 7:3, which was carried out using a Hitachi-TM3000 microscope equipped with a Bruker Quantax 70 EDS system. The crystal quality was also confirmed by x-ray Laue diffraction. The single crystals were well oriented within an error of less than 1$^\circ$ for the study of anisotropic magnetic properties.

The magnetic properties were measured with a Magnetic Property Measurement System (Quantum Design, QD) in the temperature interval  2 K\,$\leq$T $\leq$320\,K. Specific heat was measured with a QD Physical Property Measurement System (PPMS) in the temperature interval 1.90\,K\,$\leq$\,x\,$\leq$\,250\,K.  Thermal conductivity was measured from 1.9\,K to 300\,K using the Thermal Transport Option (TTO) from QD and a 14 Tesla PPMS. A rectangular bar with the dimension of 0.70\,mm\,$\times$\,0.75\,mm\,$\times$\,8\,mm was cut from a large crystal and used for the TTO measurement. Silver epoxy (H20E Epo-Tek) was utilized to provide mechanical and thermal contacts during the thermal transport measurements. Thermal conductivity measurement was performed along a random orientation because the heat transport in \textit{R}VO$_3$ perovskites shows no directional dependence.\cite{yan2004unusually,yan2007orbital}

The high-resolution x-ray single crystal diffraction was performed at 11ID-C, Advanced Photon Source, Argonne National Laboratory. Single crystal diffraction measurements were performed at 90\,K$<$T$_N$, T$_N<$130\,K$<$T*, and T$^*<$300\,K, respectively.

Part of the crystals were ground into fine powder for neutron powder diffraction measurements. Neutron powder diffraction was performed in the temperature range 12\,K\,$\leq$\,T\,$\leq$\,300\,K at HIPD, Los Alamos Neutron Science Center, Los Alamos National Laboratory.

Single crystal neutron diffraction experiments were performed using HB-1A triple-axis spectrometer, HB-3A four-circle single crystal diffractometer, and the wide angle neutron diffractometer (WAND) located at the High Flux Isotope Reactor (HFIR), and the elastic diffuse scattering spectrometer (Corelli) at Spallation Neutron Source (SNS) of the Oak Ridge National Laboratory. The HB-1A spectrometer operates with a fixed Ei=14.6 meV employing a double pyrolitic graphite (PG) monochromator system. Two PG filters were mounted after each monochromator to significantly reduce higher order contaminations of the incident beam. The WAND diffractometer uses a vertically focused Ge (1 1 3) monochromator that provides a wavelength of 1.48${\AA}$. The Y$_{0.7}$La$_{0.3}$VO$_3$ single crystal was oriented in  [H 0 L] and [0 K L] scattering planes in the HB-1A and WAND experiments, respectively. For both experiments, the crystal was mounted on aluminum plate, sealed with helium exchange gas and cooled using a closed-cycle He refrigerator. Three patterns were collected at 4 K , 130 K and 220 K using WAND by rotating the crystal with respect to the fixed position sensitive detector that covers 125  scattering angle.

Single crystal neutron diffraction was performed to measure the nuclear structures at 5 K and 150 K and magnetic structure at 5 K, at the HB-3A four-circle diffractometer at HFIR. A neutron wavelength of 1.003\,${\AA}$  was used from a bent perfect Si-331 monochromator.\cite{chakoumakos2011four} The neutron diffraction data refinements are based on 85 magnetic and nuclear reflections with the program FULLPROF.

\section{Results}

\subsection{Magnetization, specific heat and thermal conductivity}

\begin{figure} \centering \includegraphics [width = 0.47\textwidth] {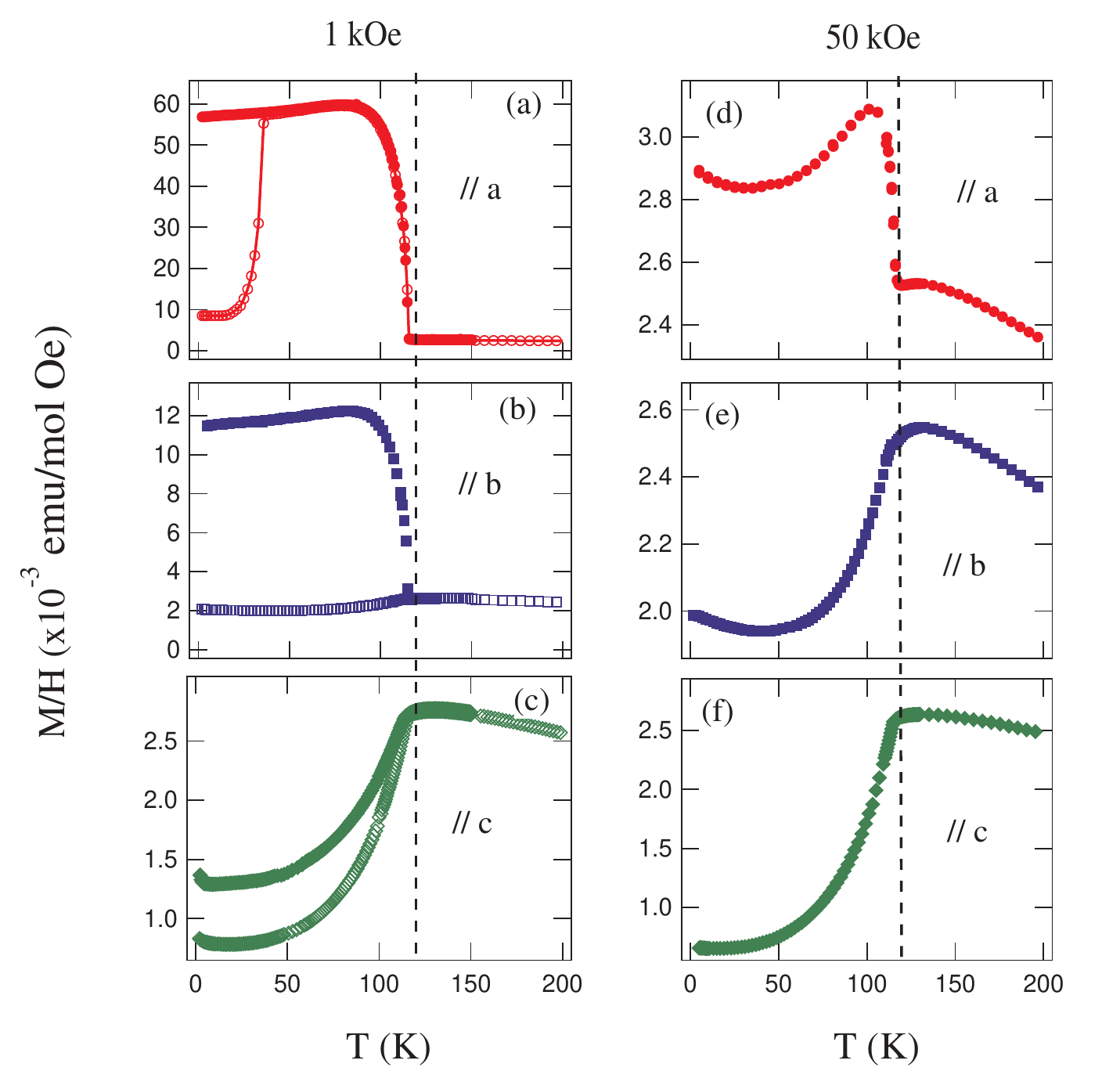}
\caption{(color online) Temperature dependence of magnetization measured in both zero-field-cooled (ZFC, open symbol) and field-cooled (FC, solid symbol) modes in an applied magnetic field of 1\,kOe and 50\,kOe along three different crystallographic axes of Y$_{0.7}$La$_{0.3}$VO$_3$. The anomaly around 40\,K in ZFC curve in (a) comes from domain alignment. }
\label{MT-1}
\end{figure}

The magnetization curves measured in zero-field-cooling (ZFC) mode are sensitive to a small remnant field of even only a few Oe. We thus quenched the magnet at 150\,K before cooling to 2\,K. Figure\,\ref{MT-1} shows the temperature dependence of the magnetization taken along three different crystallographic axes in the temperature range 2\,K$\leq$T$\leq$300\,K and in an applied magnetic field of 1\,kOe and 50\,kOe, respectively. No anomalous diamagnetism, as in LaVO$_3$,\cite{mahajan1992magnetic,tung2008spin,nguyen1995magnetic} or magnetization reversal, as in YVO$_3$,\cite{ren1998temperature,tung2007magnetization,belik2013fresh} was observed in the whole temperature range. One magnetic transition at 116\,K could be well resolved in all the curves measured with fields along different crystallographic directions. The temperature dependence of the magnetization suggests an antiferromagnetic order at T$_N$\,=\,116\,K. A weak canted-spin ferromagnetic component was observed along the \textit{a} axis below T$_N$. The canted-spin ferromagnetic component along \textit{b} axis is much weaker and can be observed in a small magnetic field. This weak canted ferromagnetism might be from the misalignment of crystals during measurement which picks up a small contribution from \textit{a}-axis.

\begin{figure} \centering \includegraphics [width = 0.47\textwidth] {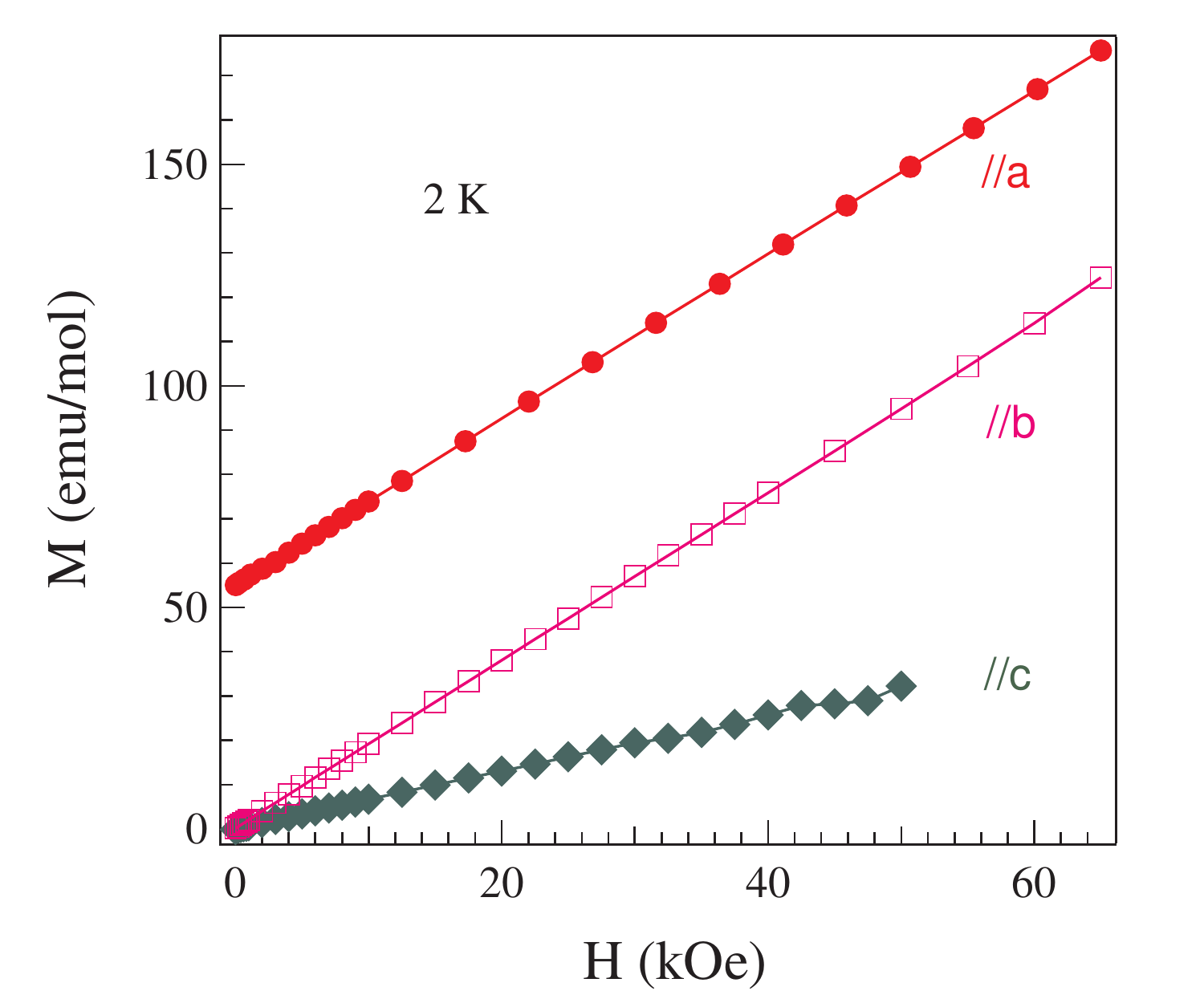}
\caption{(color online) M(H) curves measured at 2\,K along three different crystallographic axes of Y$_{0.7}$La$_{0.3}$VO$_3$.}
\label{MH-1}
\end{figure}

Figure\,\ref{MH-1} shows the M(H) curves measured at 2\,K along three crystallographic axes. A linear field dependence was observed when the external magnetic field is applied along all three crystallographic axes. However, a remanent magnetization of $\sim$50\,emu/mol was observed when the external magnetic field is applied along \textit{a}-axis. This is consistent with the observation of a ferromagnetic component along \textit{a}-axis in Fig.\ref{MT-1}.

\begin{figure} \centering \includegraphics [width = 0.47\textwidth] {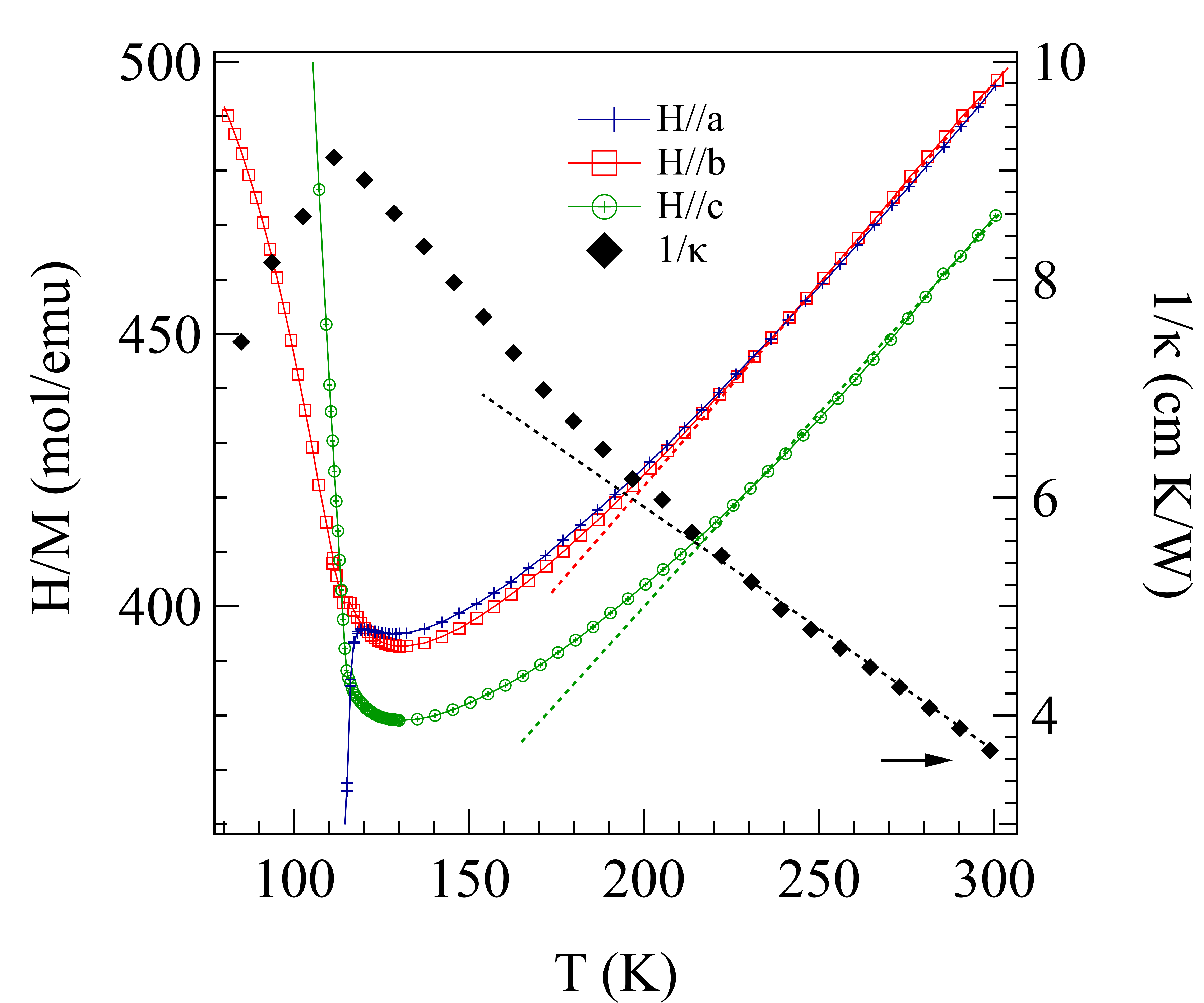}
\caption{(color online) Temperature dependence of reciprocal magnetization and thermal conductivity of Y$_{0.7}$La$_{0.3}$VO$_3$. The dashed lines highlight the slope change around 220\,K below which spin-orbital entanglement occurs.}
\label{HMK-1}
\end{figure}

Thermal conductivity (Fig.\,\ref{HMK-1}) of Y$_{0.7}$La$_{0.3}$VO$_3$ shows a similar temperature dependence as those of other Y$_{1-x}$La$_{x}$VO$_3$ (x=0.40, 0.74, 0.86, and 1) members.\cite{yan2011spin,yan2004unusually} A phonon glass behavior above T$_N$ signals strong phonon scattering by spin and orbital fluctuations. The recovery of a regular phonon behavior of thermal conductivity below T$_N$ suggests both spin and orbital ordering across T$_N$. A slope change in the temperature dependence of reciprocal thermal conductivity is observed around T*=220\,K, where the H/M curves deviate from the linear temperature dependence. This is highlighted in Fig.\,\ref{HMK-1}. As reported before,\cite{zhou2008frustrated,yan2011spin} the slope change signals the transition from orbitally disordered paramagnetic state (labelled as Disordered State I in Ref.[\citenum{yan2011spin}]) to the spin-orbital entangled state (labelled as Disordered State II in Ref.[\citenum{yan2011spin}]).

\begin{figure} \centering \includegraphics [width = 0.47\textwidth] {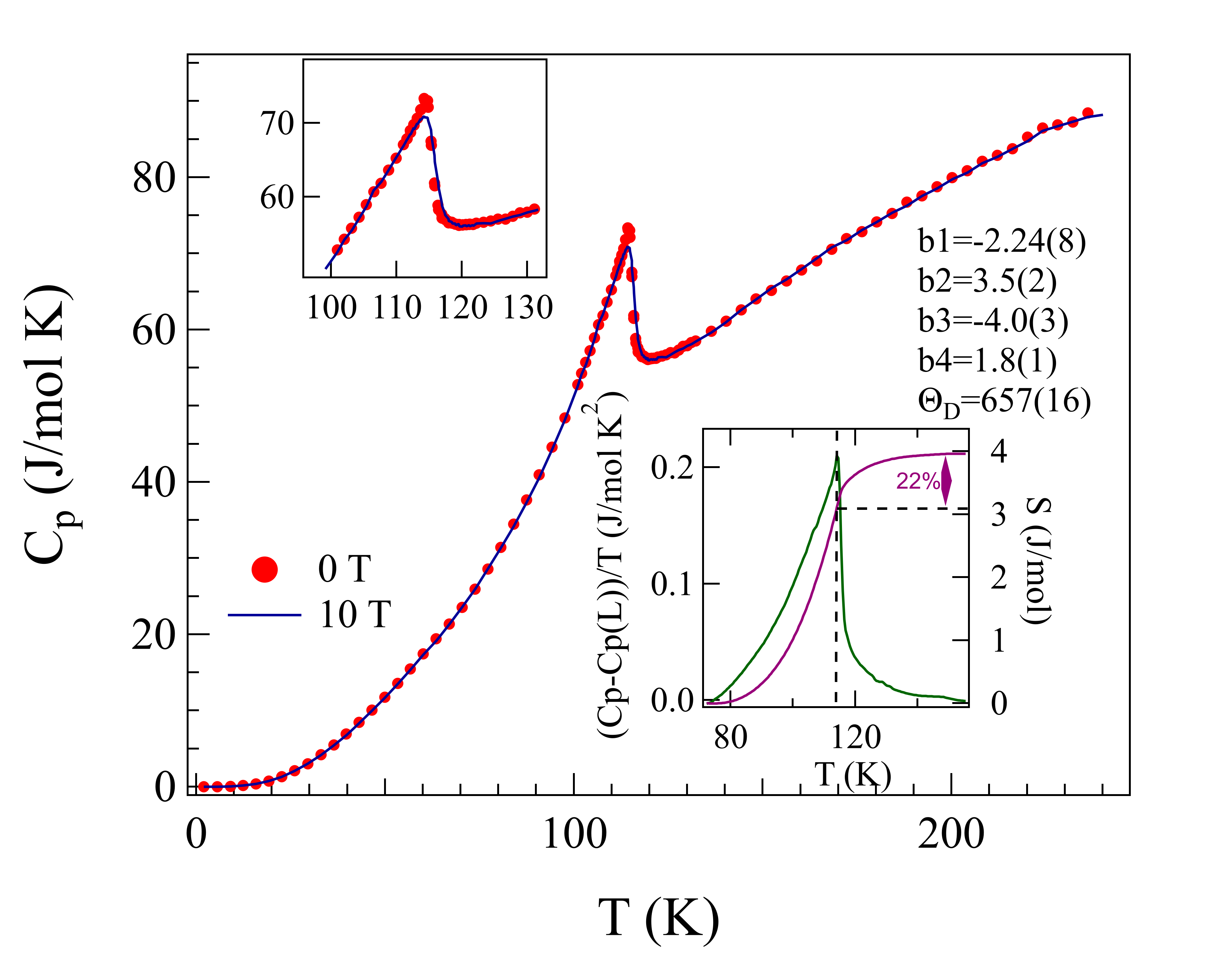}
\caption{(color online) Temperature dependence of specific heat of Y$_{0.7}$La$_{0.3}$VO$_3$ single crystal. The upper inset highlights the details around T$_N$. The lower inset shows the entropy change across T$_N$. The parameters estimating the lattice specific heat using the Thirring model are also listed in the figure. b$_i$ (i=1-4) is the fitting parameter in Thirring model\cite{thirring1913theorie}: C$_{lattice}$ = 3NR(1+b$_i\mu^{-n}$), where N is the number of atoms in the unit cell, R is the ideal gas constant, $\mu$ = (2$\pi$T/$\theta_D)^2$+1, and $\theta_D$ is the Debye temperature.}
\label{Cp-1}
\end{figure}

Figure\,\ref{Cp-1} shows the temperature dependence of specific heat measured in the magnetic fields of 0 and 100 kOe. Only one lambda-type anomaly was observed around 116 K where the magnetic order has been determined from the magnetic measurements. This lambda-type anomaly broadens and is shifted slightly to higher temperatures in an external magnetic field of 100\,kOe. The entropy release associated with orbital and/or spin ordering across T$_N$ was calculated after subtracting the lattice specific heat estimated using the Thirring model\cite{thirring1913theorie}. A total entropy change of 3.76\,J/mol K across the transition is similar to that of Y$_{0.50}$La$_{0.50}$VO$_3$ and larger than the total entropy change at T$_{OO}$ and T$_N$ for YVO$_3$.\citep{yan2011spin,blake2002neutron} If we define T$_N$ as the temperature where the maximum in the specific-heat curve is located, 22\% entropy is released above T$_N$ for Y$_{0.7}$La$_{0.3}$VO$_3$, which is less than that in Y$_{0.50}$La$_{0.50}$VO$_3$ where the largest variance is expected. The comparison with Y$_{0.50}$La$_{0.50}$VO$_3$ and YVO$_3$ suggests that both orbital and magnetic ordering might take place at T$_N$=116\,K. The lambda-type anomaly suggests that the transition at T$_N$ is of second-order type.

\subsection{Synchrotron and neutron diffraction}

In the G-type orbitally ordered state, \textit{R}VO$_3$ perovskites have a monoclinic structure (\textit{P}2$_1$/\textit{a}).\cite{blake2002neutron} Compared to the orthorhombic (\textit{Pbnm}) phase of the orbitally disordered state I and the low temperature C-OO/G-AF state, the monoclinic distortion is small and the new reflections forbidden by \textit{Pbnm} are very weak. The weak reflections cannot be resolved by a conventional laboratory x-ray source or neutron powder diffraction. \cite{blake2002neutron} We thus performed  synchrotron single crystal diffraction measurements in a high resolution mode. The measurements were performed at three temperatures: 298 K in disordered state I, 130 K in disordered state II, and 90 K below T$_N$. Our careful search did not find any weak reflections, such as (401), violating the \textit{Pbnm} symmetry. The synchrotron diffraction experiments provide strong evidence for the orthorhombic symmetry for Y$_{0.7}$La$_{0.3}$VO$_3$ in all three different temperatures. Thus, the magnetic order below T$_N$ is expected to be G-AF, which is confirmed by neutron diffraction studies presented below.

Figure\,\ref{101-1}(a) shows the temperature dependence of (101) reflection measured on a piece of single crystal at HB1-A, HFIR. This reflection is allowed in \textit{Pbnm} symmetry and has finite intensity in the paramagnetic states above T$_N$. Upon cooling across T$_N$, the intensity increases. The transition temperature determined from this order parameter measurement is 116\,K, consistent with that determined from magnetic susceptibility and specific heat measurements. It is interesting to note that (100) reflection is also observed below T$_N$. The (100) reflection is forbidden to \textit{Pbnm} and it is absent in synchrotron single crystal diffraction measurement; both point to a magnetic origin of (100) peak. The intensity of this peak is about 1\% of (101) reflection and increases below 116\,K upon cooling. To further determine the magnetic structure, neutron single crystal diffraction study was performed at 5\,K using the four-circle diffractometer at the HFIR, ORNL. Representation analysis were performed by BasIreps\cite{rodriguez1993recent} and SARAh\cite{wills2000new}. Four magnetic symmetries \textit{Pbnm}, \textit{Pb'nm'}, \textit{Pb'n'm}, and \textit{Pbn'm'} are allowed by the structure symmetry \textit{Pbnm}. Only the magnetic symmetry \textit{Pbn'm'} fits the neutron diffraction data, which is a canted antiferromagnetic structure (see Fig.\,\ref{101-1}). The R-factor is 0.0378 for 73 reflections at 5 K from the FULLPROF refinement.  The refinements are based on combined magnetic and nuclear phases. The moment is m=[0.32(18)   0.38(11)   1.40(6)]$\mu_B$ with a size of 1.49(8)\,$\mu_B$. Figure\,\ref{101-1} (b) and (c) show the moment alignment in \textit{ac} and \textit{bc} plane, respectively. The canting along \textit{a}-axis is about 14(6)$^\circ$ which gives a ferromagnetic component along \textit{a}-axis. The canting along \textit{b}-axis is about 12(8)$^\circ$ and the canted moments cancel each other out. The nuclear structure refinements were performed based on 73 and 64 reflections at 5 K and 150 K (the R-factors are 4.3\% and 6.2\%), respectively. The V-O bond length and V-O-V bond angle obtained from the refinements agree with the results from neutron powder diffraction measurements presented later.

We also performed neutron single crystal diffraction using the wide angle neutron diffractometer (WAND) located at HFIR and the elastic diffuse scattering spectrometer (Corelli) at SNS to look for possible magnetic or nuclear diffuse scattering in the temperature range T$_N\leq$T$\leq$T*. No sign of any diffuse scattering was observed at 130\,K.

\begin{figure} \centering \includegraphics [width = 0.47\textwidth] {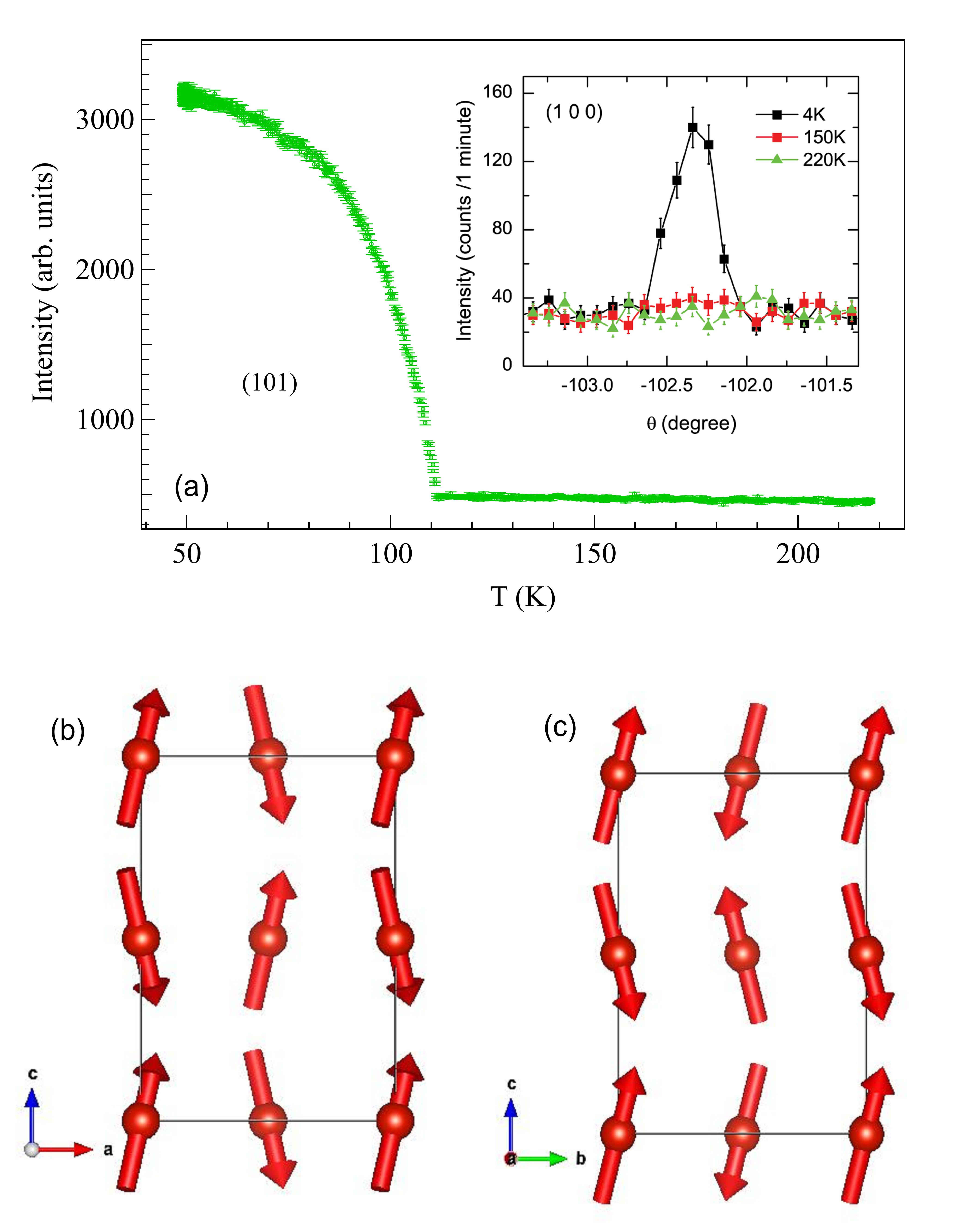}
\caption{(color online) (a) Temperature dependence of the integrated intensity of (101) reflection in the range 50\,K$\leq$T$\leq$220\,K. The inset shows the weak (100) magnetic reflection. (b) The spin alignment in \textit{ac} plane. (c) The spin alignment in \textit{bc} plane.}
\label{101-1}
\end{figure}

\begin{figure} \centering \includegraphics [width = 0.47\textwidth] {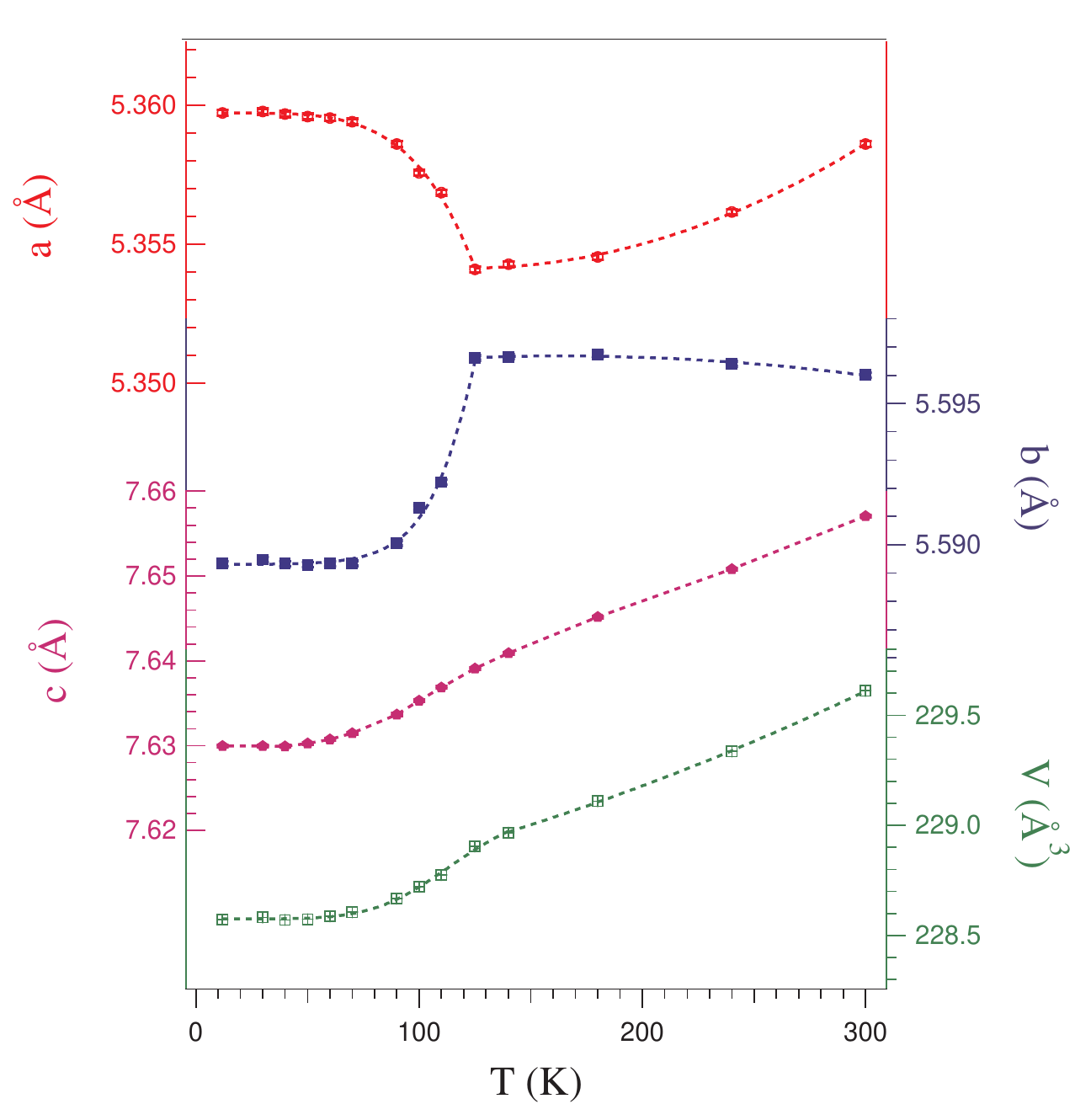}
\caption{(color online) Temperature dependence of lattice parameters obtained from Rietveld refinement of neutron powder diffraction patterns.}
\label{LP-1}
\end{figure}

Neutron powder diffraction was performed on pulverized single crystals in the temperature range 10\,K$\leq$T$\leq$300\,K to study the lattice response to the spin-orbital entanglement and magnetic order. Since no symmetry lowering was observed by synchrotron single crystal diffraction measurements, all neutron powder diffraction patterns were refined with the orthorhombic symmetry (space group \textit{Pbnm}) using GSAS.\cite{toby2001expgui}  Fig.\,\ref{LP-1} shows the temperature dependence of lattice parameters which shows a clear anomaly across T$_N$. This temperature dependence agrees with that from powder diffraction using synchrotron\cite{yan2011spin} and lab x-ray results(not shown).  The gradual change of \textit{c}-lattice parameter and the unit cell volume across T$_N$ is in contrast to the more abrupt change of \textit{a}- and \textit{b}-lattice parameters. Cooling across T$_N$, \textit{a}-lattice shows an abrupt increase while  \textit{b}-lattice an abrupt drop. Below 80\,K, all lattice parameters show little temperature dependence.

While \textit{c}-lattice and the unit cell volume show a nearly linear temperature dependence above T$_N$,  the temperature dependence of \textit{a}- and \textit{b}-lattice parameters shows a nonlinear temperature dependence. Above T*, \textit{a}-lattice  decreases and \textit{b}-lattice increases with decreasing temperature. In the temperature range T$_N<T<$T*, both \textit{a}- and \textit{b}-lattice show little temperature dependence.

\begin{figure} \centering \includegraphics [width = 0.47\textwidth] {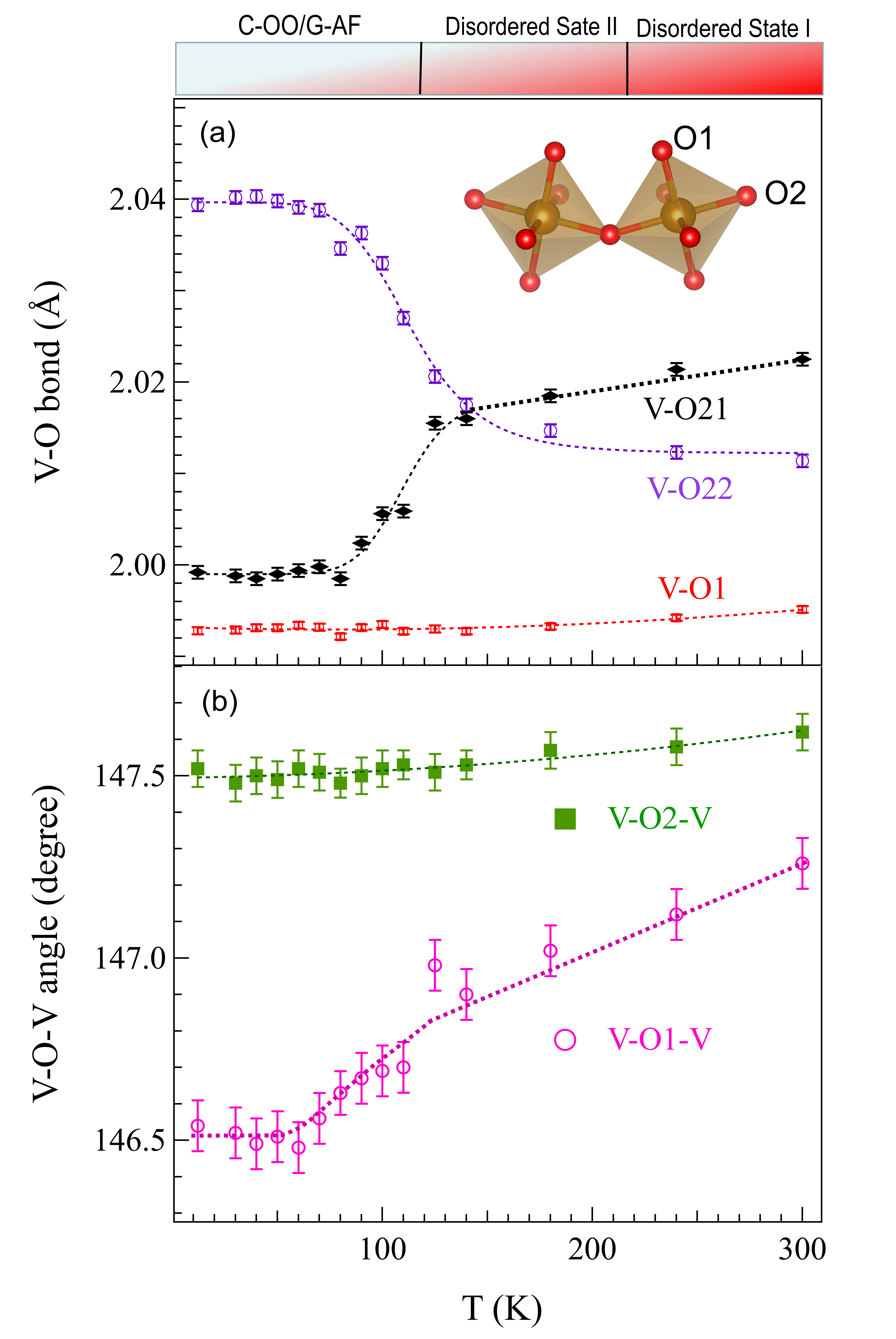}
\caption{(color online) Temperature dependence of V-O bond lengths and V-O-V bond angles obtained from Rietveld refinement of neutron powder diffraction patterns. The dashed curves are guide for the eyes.}
\label{BLBA-1}
\end{figure}

The V-O bond lengths and V-O-V bond angles obtained from the neutron powder diffraction refinements are plotted as a function of temperature and shown in Fig.\,\ref{BLBA-1}. The out-of-plane V-O1 bond is always the shortest and shows a small temperature dependence in the whole temperature range. This is consistent with the picture that xy orbital is always occupied when considering the orbital occupation below room temperature. At room temperature, the V-O1 bond is 1.9951(2)\,${\AA}$, shorter than the two in-plane V-O bonds: V-O21 (along \textit{x} direction) of 2.0114(2)\,${\AA}$ and V-O22 (along \textit{y} direction) of 2.0225(2)\,${\AA}$. These bond lengths are comparable to those of YVO$_3$.\cite{blake2002neutron,reehuis2006neutron} Upon cooling from room temperature, V-O21 decreases while V-O22 increases and the difference between V-O21 and V-O22 becomes smaller. Cooling across T$_N$, V-O21 shows a rapid drop and V-O22 an increase to a saturation value below 80\,K. At 12\,K, V-O21 (1.9992(2)\,${\AA}$) is comparable to V-O1 (1.9928(2)\,${\AA}$), V-O22 becomes the longest with a bond length of 2.0394(2)\,${\AA}$. This bond length splitting is similar to that in YVO$_3$ below T$_{CG}$.\cite{blake2002neutron,reehuis2006neutron}

Fig.\,\ref{BLBA-1}\,(b) shows the temperature dependence of V-O-V bond angles. In the whole temperature range, the in-plane V-O2-V bond angle is larger than the out-of-plane V-O1-V. At room temperature, the V-O-V bond angles $\sim$147$^\circ$ are about 3$^\circ$ larger than those of $\sim$144$^\circ$ in YVO$_3$.\cite{blake2002neutron,reehuis2006neutron} This is consistent with the facts that La$^{3+}$ has a larger ionic radius than Y$^{3+}$ and the partial substitution of Y$^{3+}$  by La$^{3+}$ increases the average ionic radius of the rare earth site. Above T$_N$, the in-plane V-O2-V bond angle decreases at a rate of 5.8(8)$\times$10$^{-4}$ degree per Kelvin with decreasing temperature; below T$_N$, it shows little temperature dependence. The out-of-plane V-O1-V bond angle decreases while cooling at a much faster rate of 2.2(2)$\times$10$^{-3}$ degree per Kelvin above T$_N$; a large drop of $\sim$0.3\,degree is observed cooling across T$_N$; below 80\,K, V-O1-V bond angle shows little temperature dependence.

\section{Discussion}
Our results support a ground state of C-OO and canted G-AF below T$_N$=116\,K for  Y$_{0.7}$La$_{0.3}$VO$_3$. The entropy change (Fig.\,\ref{Cp-1}) across the magnetic transition is larger than that at T$_{OO}$ or T$_{N}$ for YVO$_3$. Since only one anomaly was observed at T$_N$ in the temperature dependence of magnetic susceptibility, specific heat, and thermal conductivity, we can conclude that spin and orbital order simultaneously at T$_N$ in Y$_{0.7}$La$_{0.3}$VO$_3$. The synchrotron and neutron diffraction studies show that the orthorhombic symmetry is maintained in the whole temperature range investigated in this work. The G-type magnetic ground state and the orthorhombic symmetry rule out G-OO below T$_N$ which is compatible with C-AF and a monoclinic lattice. The C-OO state below T$_N$ is also supported by the octahedral distortion and the observed G-type AF order. The V-O bond-lengths below T$_N$ determined by neutron powder diffraction (Fig.\,\ref{BLBA-1}) are similar to those in YVO$_3$ below 77\,K with VO22$>$VO21$\approx$VO1.\cite{blake2002neutron,reehuis2006neutron} The neutron single crystal diffraction confirmed the G-type antiferromagnetic structure below T$_N$ with the moment direction primarily along c-axis. In the orthorhombic \textit{Pbnm} structure, the Dzyaloshinskii vector \textbf{D}$_{ij}$ is parallel to the \textit{b}-axis. Thus in the C-OO state below T$_N$, both the antisymmetric exchange term \textbf{D}$_{ij}\cdot\textbf{S}_i\times\textbf{S}_j$ and a site anisotropy give a ferromagnetic canted-spin component parallel to the \textit{a}-axis. This is confirmed by the magnetic measurements (Fig. \ref{MT-1}, \ref{MH-1}).

In the paramagnetic state, both the reciprocal magnetic susceptibility and reciprocal thermal conductivity show a slope change (see Fig. \ref{HMK-1}) around 220\,K. This slope change was observed in a wide compositional range 0.2$<$x$\leq$1.0 of Y$_{1-x}$La$_x$VO$_3$ and signals the spin-orbital entanglement in the temperature range $T_N\leq$T$\leq$T*.\cite{zhou2008frustrated,yan2011spin} Therefore, the local lattice distortion of Y$_{0.7}$La$_{0.3}$VO$_3$ in this temperature range can be a general feature for the spin-orbital entangled state of Y$_{1-x}$La$_x$VO$_3$ (0.2$<$x$\leq$1.0).

The most striking feature of Fig.\,\ref{BLBA-1} is that the difference between in-plane V-O21 and V-O22 bonds becomes smaller upon cooling from room temperature and they are comparable in the spin-orbital entangled state. This local structural distortion is incompatible with the octahedral-site distortion intrinsic to the \textit{Pbnm} perovskite compounds and also different from that in the orbitally ordered state.  The universal features of lattice distortions in perovskites with orthorhombic \textit{Pbnm} space group and the biasing effect on orbital ordering in \textit{RM}O$_3$ (\textit{M}=transition metal) have been well addressed by Zhou et al.\cite{zhou2005universal,zhou2008intrinsic,zhou2010intrinsic,zhou2009orbital} In addition to the GdFeO$_3$-type distortion, which can be described by the tilting system $a^-a^-b^+$ in the Glazer notation, octahedral site distortion leads to in \textit{ab} plane one long and one short O-M-O bond alternating along the pseudocubic [100] and [010] axes. The difference between the long and short in-plane M-O bonds increases with decreasing temperature since the tolerance factor of perovskites have a positive temperature dependence.  At room temperature, the VO$_{6/2}$ octahedra in Y$_{0.7}$La$_{0.3}$VO$_3$ have three inequivalent V-O bonds: V-O21$>$V-O22$>$V-O1, consistent with the above analysis. However, it is unusual that the difference between V-O21 and V-O22 decreases upon cooling and these two in-plane V-O bonds become comparable in the spin-orbital entangled state. Obviously, the octahedral site distortion compatible with the spin-orbital entanglement is incompatible with the intrinsic lattice distortion of \textit{Pbnm} perovskites. This incompatibility leads to structural instability responsible for the anomalous thermal conductivity in the spin-orbital entangled state and results in the phase transition from spin-orbital entangled state to the C-OO/G-AF ground state at T$_N$=116\,K for Y$_{0.7}$La$_{0.3}$VO$_3$. In the C-OO/G-AF state, the bond length relation of V-O22$>$V-O21$\approx$V-O1 is similar to that in YVO$_3$.

\begin{figure} \centering \includegraphics [width = 0.47\textwidth] {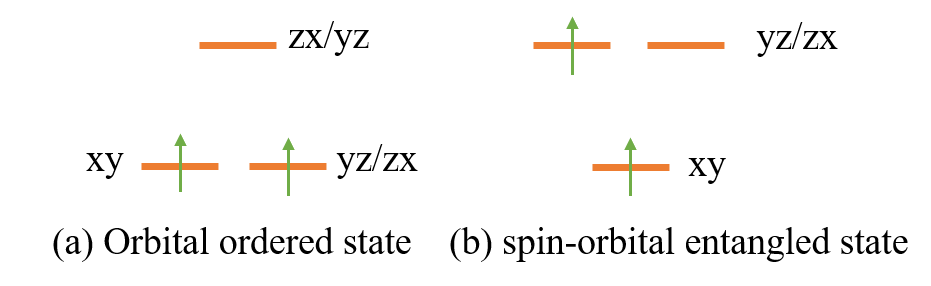}
\caption{(color online) The orbital occupancy in (a) orbitally ordered state and (b) spin-orbital entangled state.}
\label{t2g-1}
\end{figure}

Following the classic Jahn-Teller orbital physics, the distinctly different octahedral distortion results in the different orbital occupancy. In both G-OO and C-OO states, the short in-plane V-O bond is comparable to the out-of-plane V-O1. The orbital occupancy in case of this type of octahedral site distortion is illustrated in Fig.\,\ref{t2g-1}(a). The same orbital occupancy is maintained with the trigonal distortion of VO$_{6/2}$ octahedra.\cite{landron2008importance} Fig.\,\ref{t2g-1}(b) shows the proposed orbital occupancy with V-O21$\approx$V-O22$>$V-O1 as observed in Y$_{0.7}$La$_{0.3}$VO$_3$ in the temperature range T$_N\leq T \leq$T*. This type octahedral distortion results in the yz/zx orbital degeneracy, which facilitates the spin-orbital entanglement in the temperature range T$_N\leq$ T $\leq$T*.

The difference between VO21 and VO22 bonds in Y$_{0.7}$La$_{0.3}$VO$_3$ around room temperature is much larger than that in LaVO$_3$. If the bondlength difference affects the temperature below which the spin-orbital entanglement occurs, we would expect T* for LaVO$_3$ is much higher than that for Y$_{0.7}$La$_{0.3}$VO$_3$. However, we noticed that T* is around 200\,K and independent of x in Y$_{1-x}$La$_x$VO$_3$ (0.20$<$x$\leq$1). This signals that T* does not depend on the difference of the in-plane V-O bonds in Y$_{1-x}$La$_x$VO$_3$ or the average GdFeO$_3$ distortion. It is worth mentioning that for \textit{R}VO$_3$ members with \textit{R}=Lu,...,Pr, T$_{OO}$, where the G-type orbital order occurs, is also around 200\,K. Similar T$_{OO}$ and T* suggests that the spin-orbital entangled state and G-OO state are comparable in energy. This puts some constraint on the theoretical study of the spin-orbital entanglement in \textit{R}VO$_3$ perovskites.

The competition between the G-OO state and the spin-orbital entangled state in \textit{R}VO$_3$ perovskites is biased by the quenched disorder at the rare earth site. Y$_{0.7}$La$_{0.3}$VO$_3$ and EuVO$_3$ have similar average ionic radius of the rare earth site. EuVO$_3$ has a G-OO/C-AF ground state with T$_{OO}$=204\,K and T$_N$=131\,K.\cite{tung2007heat,fukuta2011effects} The occurrence of spin-orbital entanglement and the C-OO/G-AF ground state of Y$_{0.7}$La$_{0.3}$VO$_3$ highlight the importance of the size variance of the rare earth site: rare earth site disorder favors the spin-orbital entanglement by impeding the cooperative Jahn-Teller distortion. This is supported by the observation of the spin-orbital entanglement in a wide composition range 0.2$<$x$<$1 in Y$_{1-x}$La$_x$VO$_3$. The pseudocubic LaVO$_3$ is special where the spin-orbital entanglement wins against G-OO without the help of quenched disorder. For CeVO$_3$, although all studies suggest the G-OO/C-AF ground state, the sequence of spin and orbital ordering is different. Several groups reported a T$_N$=136\,K which is followed by a first-order structural transition at T$_t$=131\,K. \cite{miyasaka2003spin,yan2004unusually,munoz2003crystallographic,nguyen1995magnetic} This sequence of spin and orbital ordering is similar to that in LaVO$_3$. However, Ren et al and Reehuis et al  observed an orbital order at T$_{OO}$=154\,K (136K) and a magnetic order at T$_{N}$=136\,K(124).\cite{ren2003orbital,reehuis2008crystal} The discrepancy signals that the spin and orbital ordering in CeVO$_3$ is sensitive to the stoichiometry which induces disorder and affects the tolerance factor.

The spin-orbital entanglement is suppressed in the orbitally ordered state. Considering a weak orbital-lattice coupling for t$_{2g}$ electrons, the quantum fluctuations of spin-orbital entangled phase might compete with the G-OO phase below T$_{OO}$ for \textit{R}VO$_3$ members with \textit{R}=Lu,...,Pr. Since the orbital flipping transition between G-OO/C-AF and C-OO/G-AF states at T$_{CG}$ only shows up in \textit{R}VO$_3$ members with \textit{R}=Lu,...,Dy, it is necessary to take into account the covalency between rare-earth-site cations and oxygens\cite{mizokawa1999interplay} and/or the e-orbital occupation due to a hybridization between $t^2$ and \textit{et} configurations.\cite{yan2007orbital,zhou2009orbital} The intrasite hybridization, which depends on the trigonal distortion,\cite{landron2008importance} are active for these \textit{R}VO$_3$ members. The effects of all these factors on the orbital flipping transition at T$_{CG}$ deserve further experimental and theoretical studies.

One important indication of the results presented in this work is about the nature of transitions at T$_N$ and T$_t$ in LaVO$_3$. LaVO$_3$ shows similar anomalies in the temperature dependence of magnetic susceptibility and thermal conductivity as other Y$_{1-x}$La$_{x}$VO$_3$ (0.20$<x<$1) members. We thus expect similar octahedral site distortions in the spin-orbital entangled state. The magnetic order at T$_N$ for LaVO$_3$ does not result from the magnetic coupling in the paramagnetic state but corresponds to the transition from the spin-orbital entangled state to the C-OO/G-AF state. As in Y$_{0.7}$La$_{0.3}$VO$_3$, the structure is still orthorhombic in the temperature interval T$_t<T<$T$_N$ for LaVO$_3$. At T$_t$, C-OO/G-AF is not stable and changes to the G-OO/C-AF which is the ground state for \textit{R}VO$_3$ with large rare earth ions.\cite{mizokawa1999interplay} This orbital flipping transition is a first order transition and similar to what happens in DyVO$_3$ at low temperatures.\cite{miyasaka2007magnetic,yan2013dy,zhou2009orbital}

\section{Summary}

In summary, we investigate the lattice response to the spin orbital entanglement in Y$_{0.7}$La$_{0.3}$VO$_3$. In the spin-orbital entangled state, the VO$_{6/2}$ octahedra have two comparable in-plane V-O bonds which are longer than the out-of-plane VO1 bond. This local structural distortion  supports the spin-orbital entanglement with degenerate yz/zx orbitals that are partially filled. However, this distortion is incompatible with the octahedral-site distortion intrinsic to the orthorhombic perovskites; and their competition induces a second order transition from the spin-orbital entangled state to the C-OO/G-AF state. The fact that both T*, below which spin-orbital entanglement occurs in Y$_{1-x}$La$_{x}$VO$_3$ (0.20$<x\leq$1), and T$_{OO}$, below which orbital orders into G-OO in \textit{R}VO$_3$ (\textit{R}=Pr, ..., Lu, and Y), are around 200\,K suggests that these two states are comparable in energy and compete with each other. Their interplay and competition lead to the rich phase diagrams of spin and orbital ordering in Y$_{1-x}$La$_{x}$VO$_3$ and \textit{R}VO$_3$. The quenched disorder at the rare earth site favors the spin-orbital entangled state instead of the cooperative Jahn-Teller distortions.  This study indicates a C-OO/G-AF state in the temperature interval T$_t$=141\,K$\leq$T$\leq$T$_N$  and an orbital flipping transition at T$_t$ for LaVO$_3$.

\section{Acknowledgments}

The authors thank Dr. S. Chang for his exploratory neutron diffraction work on this compound at the early stage of this project. Work at ORNL was supported by the U.S. Department of Energy, Office of Science, Basic Energy Sciences, Division of Materials Sciences and Engineering (JQY and MAM). Ames Laboratory is operated for the US Department of Energy by Iowa State University under Contract No. DE-AC02-07CH11358 (RJM). QC acknowledges the support of the Center for Emergent Materials, an NSF MRSEC, under Award Number DMR-1420451. JGC is supported by the MOST, NSFC, and CAS (Grant Nos. 2018YFA0305700, 11574377, XDB07020100 and QYZDB-SSW-SLH013). JSZ acknowledges the support from Gordon and Betty Moore Foundation EPiQS Initiative through a sub-contract to Grant No. GBMF4534. A portion of this research used resources at the High Flux Isotope Reactor (and/or Spallation Neutron Source, as appropriate), a DOE Office of Science User Facility operated by the Oak Ridge National Laboratory. Work at Argonne National Laboratory was supported by the US Department of Energy, Division of Basic Energy Science, under Contract No. DE-AC02-CH11357. This work has benefited from the use of HIPD at the Lujan Center at Los Alamos Neutron Science Center, funded by DOE Office of Basic Energy Sciences.  Los Alamos National Laboratory is operated by Los Alamos National Security LLC under DOE Contract DE-AC52-06NA25396.

\section{references}

%

\end{document}